\documentclass{article}
\usepackage{spconf,amsmath,graphicx}
\usepackage{CJKutf8}
\usepackage{booktabs}
\usepackage{textcomp}
\usepackage{multirow}
\usepackage{dcolumn}
\usepackage{subfigure}
\usepackage{marvosym}
\usepackage{color}

\title{WenetSpeech: A 10000+ Hours Multi-domain Mandarin Corpus for Speech Recognition}
\name{
\begin{tabular}{c}
\it Binbin Zhang$^{1,2,4,*}$, Hang Lv$^{1,4,*}$, Pengcheng Guo$^1$, Qijie Shao$^1$, Chao Yang$^{2,4}$, Lei Xie${^1}\textsuperscript{\Letter}$, Xin Xu$^3$, Hui Bu$^3$, \\
\it Xiaoyu Chen$^2$, Chenchen Zeng$^2$, Di Wu$^{2,4}$, Zhendong Peng$^{2,4}$
\end{tabular}
}

\address{
  $^1$ Audio, Speech and Language Processing Group (ASLP@NPU),\\
  School of Computer Science, Northwestern Polytechnical University\\
  $^2$Mobvoi Inc. $^3$Beijing Shell Shell Technology Co., Ltd. $^4$WeNet Open Source Community \\
  \small{\texttt{hanglv@nwpu-aslp.org,lxie@nwpu.edu.cn}}
  \thanks{* Co-first authors, equal contribution. \Letter~Corresponding author.}
}

\begin{document}
%
\maketitle
\ninept
\begin{abstract}
In this paper, we present \textit{WenetSpeech}, a multi-domain Mandarin corpus consisting of 10000+ hours high-quality labeled speech, 2400+ hours weakly labeled speech, and about 10000 hours unlabeled speech, with 22400+ hours in total.
We collect the data from YouTube and Podcast, which covers a variety of speaking styles, scenarios, domains, topics and noisy conditions.
An optical character recognition (OCR) method is introduced to generate the audio/text segmentation candidates for the YouTube data on the corresponding video subtitles, while a high-quality ASR transcription system is used to generate audio/text pair candidates for the Podcast data.
Then we propose a novel end-to-end label error detection approach to further validate and filter the candidates.
We also provide three manually labelled high-quality test sets along with WenetSpeech for evaluation -- \textit{Dev} for cross-validation purpose in training, \textit{Test\_Net}, collected from Internet for \textit{matched} test, and \textit{Test\_Meeting}, recorded from real meetings for more challenging \textit{mismatched} test.
Baseline systems trained with WenetSpeech are provided for three popular speech recognition toolkits, namely Kaldi, ESPnet, and WeNet, and recognition results on the three test sets are also provided as benchmarks. To the best of our knowledge, WenetSpeech is the current largest open-source Mandarin speech corpus with transcriptions, which benefits research on production-level speech recognition.

\end{abstract}
\vspace{-1.5em}
\section{Introduction}
\vspace{-1em}

In the past decade, the performance of automatic speech recognition (ASR) systems have been significantly improved. On the one hand, the development of neural networks has increased the capacity of models, pushing the dominant framework from the hybrid hidden Markov models~\cite{hinton2012deep,dahl2012context} to end-to-end models, like CTC~\cite{graves2006connectionist,amodei2016deep}, RNN-T~\cite{graves2012sequence,graves2013speech,wang2021cascade,wang2021efficient}, and encoder-decoder based models~\cite{chorowski2015attention,chan2016listen,kim2017joint,dong2018speech,gulati2020conformer}. To simply implement such advanced models and obtain state-of-the-art reproducible results, researchers also release several open source toolkits, including Kaldi~\cite{povey2011kaldi}, Sphinx~\cite{lee1990overview}, Fariseq~\cite{ott2019fairseq}, ESPnet~\cite{watanabe2018espnet}, and recently WeNet~\cite{yao2021wenet}, etc. On the other hand, self-supervised speech representation learning methods are proposed to make better use of a large amount untranscribed data, such as wav2vec~\cite{schneider2019wav2vec}, wav2vec 2.0~\cite{baevski2020wav2vec}, Hubert~\cite{hsu2021hubert}, and wav2vec-U~\cite{baevski2021unsupervised}, etc. In addition to these algorithm-level efforts, the development of open source corpora is also crucial to the research community, especially for academia or small-scale research groups.

Most of the current open source speech corpora for ASR benchmark in the literature remain small size and lack of domain diversities. For example, the commonly used English speech corpus - Librispeech~\cite{panayotov2015librispeech}, which includes about 1000 hours reading speech from audiobook, currently has a word error rate (WER) of 1.9\% on its test-clean benchmark. However, industrial ASR systems are usually trained with tens of thousands of hours of data with acoustic diversity and domain coverage.
To close the gap between industrial system and academic research, we notice that several large-scale multi-domain English corpora, including The People's Speech~\cite{galvez2021the}, MLS~\cite{pratap20_interspeech} and GigaSpeech~\cite{chen21o_interspeech}, are made available recently. The representative GigaSpeech
consists of 10000 hours of high quality transcribed English speech for supervised ASR training and 40000 hours audio in total for semi-supervised or unsupervised training, contributing to the research community for developing more generalized ASR systems.
Comparing with those English corpora, the largest open source Mandarin speech data is AIShell-2~\cite{du2018aishell}, including 1000 hours speech recorded in a quiet environment and having a state-of-the-art character error rate of 5.35\%. It is too simple to do further research and ASR systems developed based on it may be susceptible to performance degradation in the complex real-world scenarios. In addition, current open source Mandarin corpora are also unable to train a well generalized pre-trained model, since both the Wav2vec 2.0 large model~\cite{baevski2020wav2vec} and the XLSR-53 model~\cite{conneau2020unsupervised} are trained based on more than 50000 hours of English speech data.

In this work, we release WenetSpeech, a large multi-domain Mandarin corpus licensed for non-commercial usage under CC-BY 4.0. ``We'' means connection and sharing, while ``net'' means all of the data are collected from the Internet which is repository for diversity. The key features of WenetSpeech include:

\vspace{-0.5em}
\begin{itemize}
\item \textbf{Large scale}. 10000+ hours labeled data, 2400+ hours weakly labeled data, and about 10000 hours unlabeled data are provided, resulting in 22400+ hours audio in total.
\vspace{-0.2em}
\item \textbf{Diverse}. The data are collected from multiple speaking styles, scenarios, domains, topics, and noisy conditions.
\vspace{-0.2em}
\item \textbf{Extensible}. An extensible metadata is designed to extend the data in the future.
\end{itemize}
\vspace{-0.5em}


To the best of our knowledge, WenetSpeech is the current largest open source Mandarin speech corpus with domain diversity to satisfy various speech recognition tasks. In Section~\ref{sec:pipeline}, we first introduce the construction procedure of WenetSpeech with a reliable pipeline to obtain high-quality transcriptions, including OCR-based caption recognition on Youtube video, ASR-based automatic transcription on Podcast audio, as well as a new end-to-end label error detection approach. The corpus composition is described in Section~\ref{sec:corpus} and baseline benchmarks built on Kaldi, ESPnet and WeNet toolkits are introduced in Section~\ref{sec:exp}. We believe our corpus will bring benefit to research community for developing more generalized ASR systems.

\vspace{-1em}
\section{Creation Pipeline}\label{sec:pipeline}
\vspace{-1em}
In this section, we introduce the detailed creation pipeline of our WenetSpeech corpus, including original audio collection, audio/text segmentation candidate generation, and candidate calibration.

\vspace{-1em}
\subsection{Stage 1: Audio Collection}
\vspace{-0.5em}
In the beginning, we manually define the domains into 10 categories, which including \textit{audiobook, commentary, documentary, drama, interview, reading, talk, variety, and others}. Then, we collected and tagged the audio files from YouTube and Podcast playlists according to our selected categories. Especially, for the YouTube data, the videos are also downloaded for preparing the audio/text segmentation candidates with our OCR system. For Podcast, since the manually transcribed Mandarin data is limited, we only considered the category information and prepared to transcribed it by a high-quality ASR system.  

\vspace{-1em}
\subsection{Stage 2: Candidates Generation}
\vspace{-0.5em}
In this part, we introduce the specific pipelines to obtain the audio/text segmentation candidates from YouTube data by an OCR system and Podcast data by a high-quality ASR system. At last, text normalization\footnote{https://github.com/speechio/chinese\_text\_normalization} technique is applied to all the candidates.

\vspace{-1em}
\subsubsection{YouTube OCR}
\vspace{-0.5em}

\begin{figure}[bh]
  \vspace{-1.5em}
  \centering
  \includegraphics[width=\linewidth]{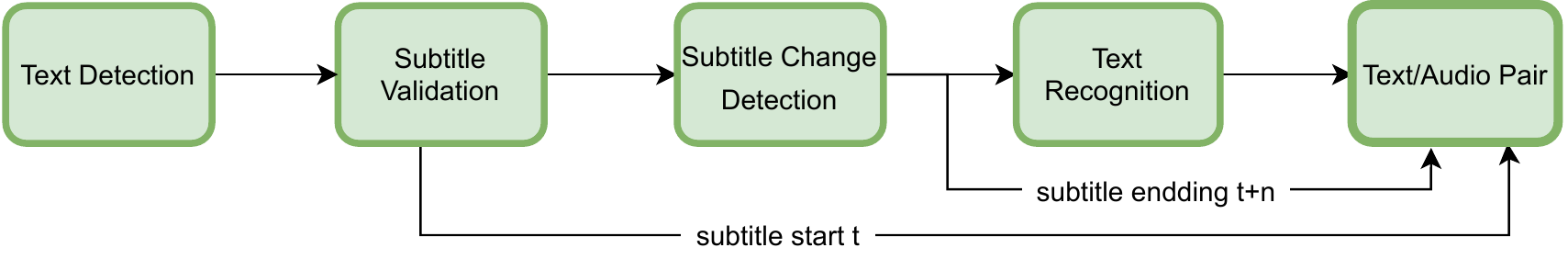}
  \vspace{-2em}
  \caption{OCR based YouTube data collection pipeline}
  \label{fig:ocr_pipeline}
  \vspace{-1em}
\end{figure}

As Figure \ref{fig:ocr_pipeline} shown, an OCR-based pipeline is applied for generating candidates from embedded subtitles on YouTube videos.

\vspace{-0.5em}
\begin{enumerate}
    \setlength{\itemsep}{0pt}
    \item \textit{Text Detection}: apply CTPN~\cite{tian2016detecting} based text detection on each frame image in the video. 
    \item \textit{Subtitle Validation}: mark frame $t$ as the \textit{start point} of a specific \textit{subtitle phrase}, when a phrase of text is detected at the bottom of the screen for subtitles at frame $t$. 
    \item \textit{Subtitle Change Detection}: compute the structural similarity (SSIM) of subtitle area frame by frame until a change is detected at frame $t+n$. Then, the frame $t+n-1$ is marked as the the \textit{end point} of this \textit{subtitle phrase}.
    \item \textit{Text Recognition}: a CRNN-CTC~\cite{shi2016end} based text recognition approach is used to recognize the detected subtitle area.
    \item \textit{Audio/Text Pair}: prepare each audio/text pair segmentation candidate with the corresponding (\textit{start point, end point, subtitle phrase}) tuple.
\end{enumerate}

To verify whether the proposed OCR-based pipeline is reliable,
we randomly extracted 5000 subtitle transcriptions from the YouTube data with different themes and manually annotated them as benchmarks by professionals for testing the \textit{Text Recognition} module.
Finally, we obtain 98\% accuracy on the test set and confirm the reliability of our OCR-based pipeline for WenetSpeech corpus.

\begin{figure}[th]
\centering
\subfigure[Audiobook]{
\includegraphics[width=0.45\linewidth]{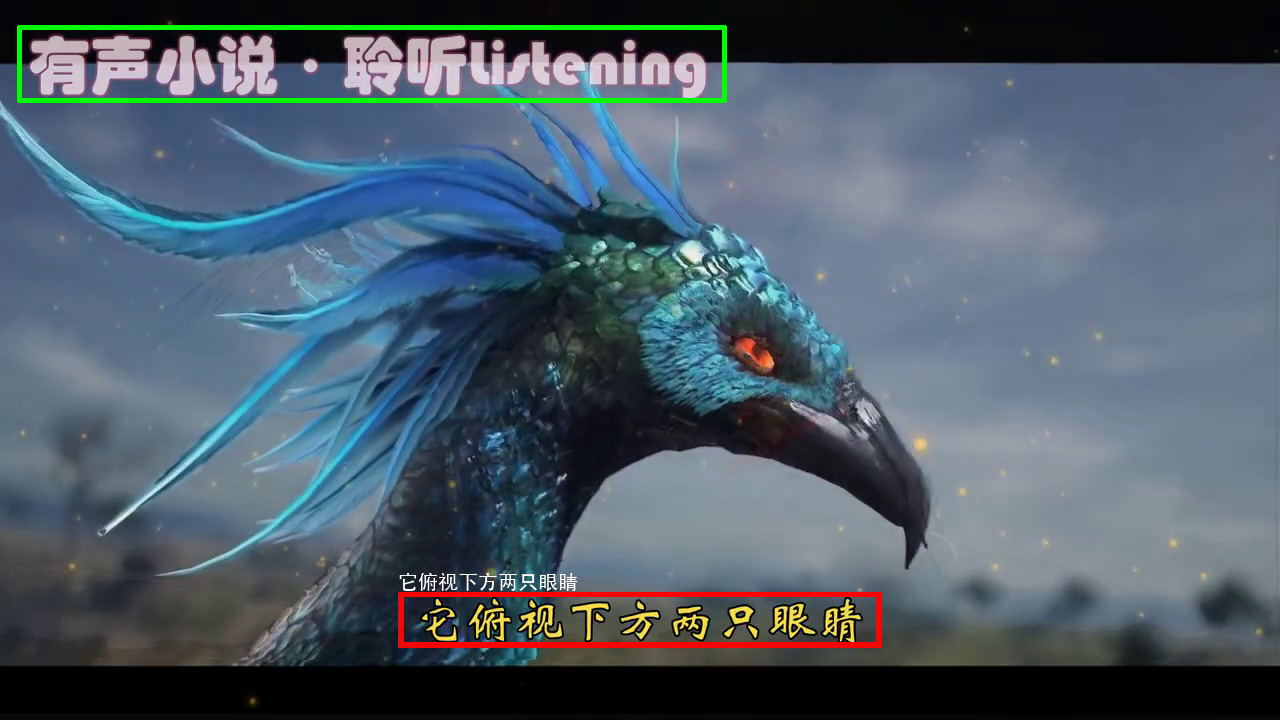}
}
\quad
\subfigure[Game commentary]{
\includegraphics[width=0.45\linewidth]{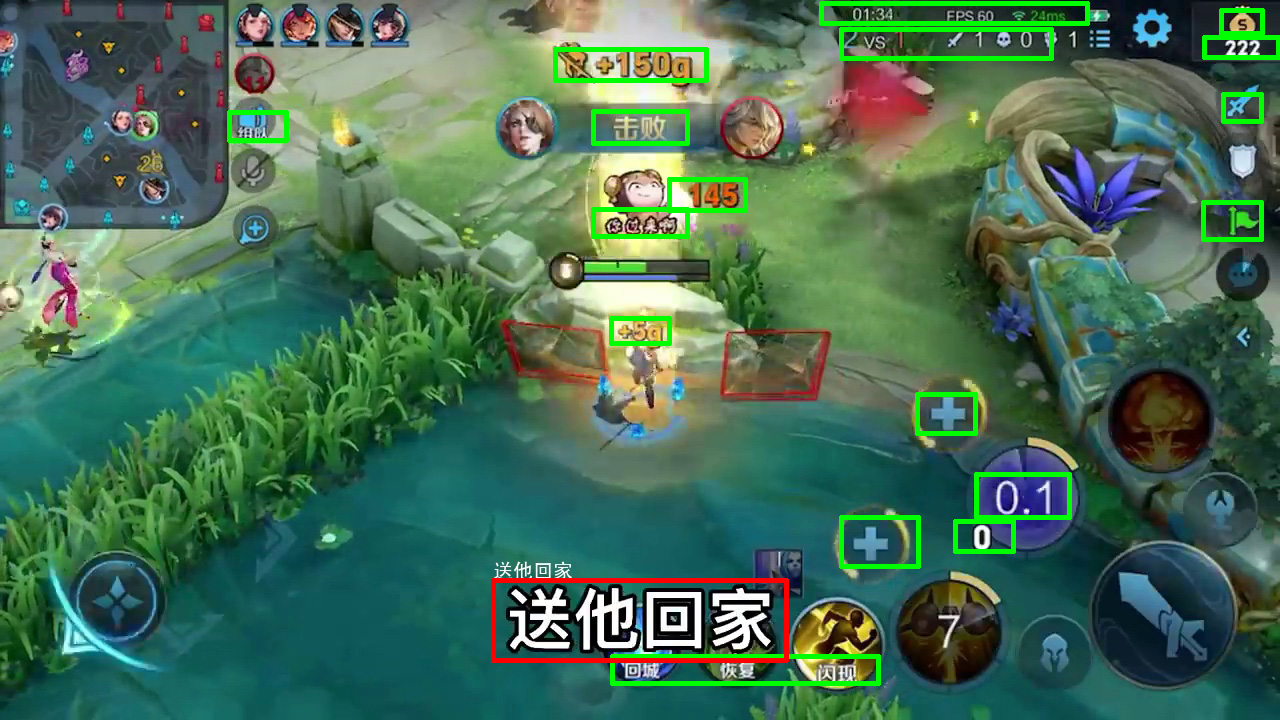}
}
\quad
\subfigure[Drama]{
\includegraphics[width=0.45\linewidth]{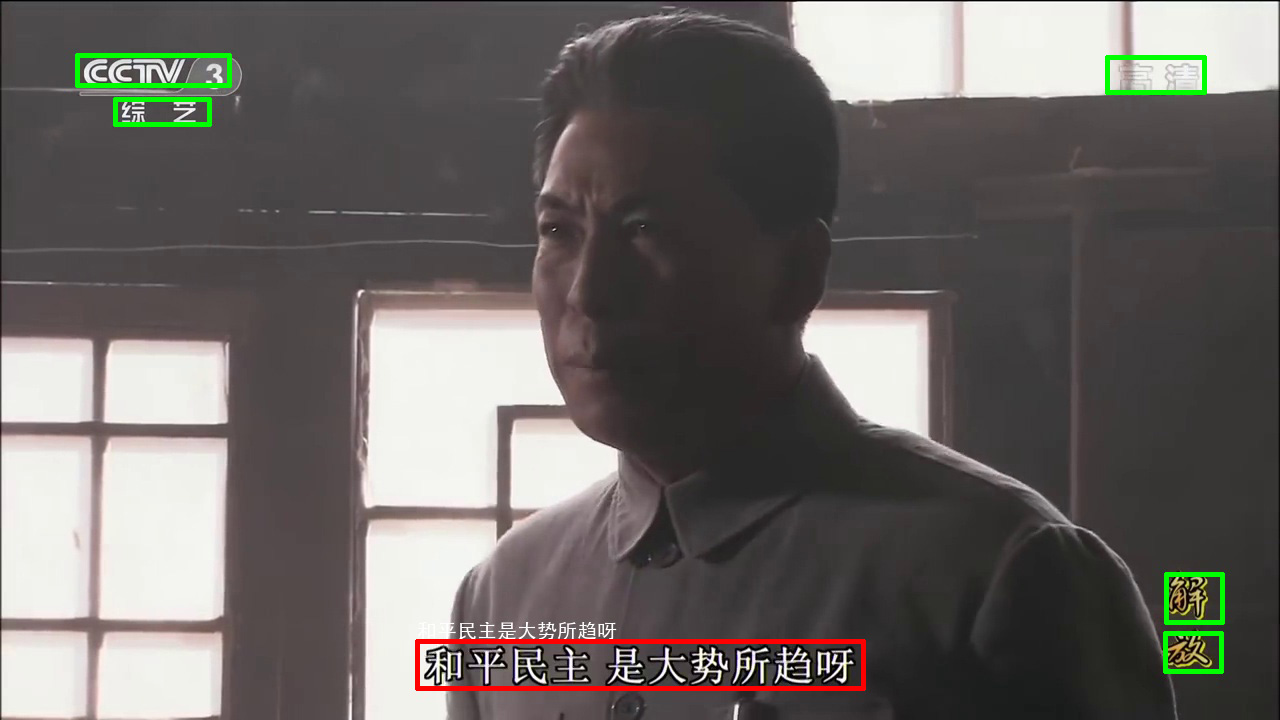}
}
\quad
\subfigure[Lecture]{
\includegraphics[width=0.45\linewidth]{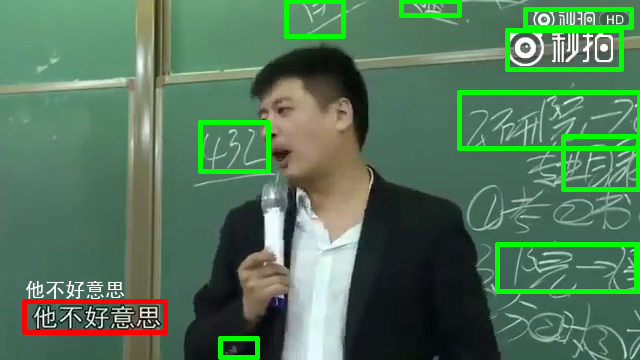}
}
\vspace{-1em}
\caption{Example outputs of the OCR pipeline}
\label{fig:ocr_example}
\vspace{-2em}
\end{figure}

Figure \ref{fig:ocr_example} shows 4 typical examples of our OCR-based system. The results of the \textit{Text Detection} module are marked with boxes. If the box is in a reasonable subtitle area, it will be marked as red, and then, the subtitle box is further processed by the \textit{Text Recognition} module. At last, the recognition result is shown above each subtitle box.

In addition, we find that some annotators prefer to split a long subtitle phrase into many pieces for a video, so that the problem of audio and subtitle asynchronous is introduced. This leads an inaccurate subtitle boundary detection problem to our OCR system. To alleviate it, we merge the consecutive video phrases until the audio is over 8 seconds.

\vspace{-1em}
\subsubsection{Podcast Transcription}
\vspace{-0.5em}
We use a third-party commercial ASR transcription system to transcribe all the Podcast data. The transcription system is one of the best system on the public benchmark platform\footnote{https://github.com/SpeechColab/Leaderboard}, and more than 95\% accuracy rates have been achieved in most of the testing scenarios, including news, reading, talk show, conversation, education, games, TV, drama, and so on.

The transcription system first segments the original audio into short segments by a VAD module, and then the audio/text pair segmentation candidates are generated by speech recognition.

\vspace{-1em}
\subsection{Stage 3: Candidates Validation}
\vspace{-0.5em}
Although the used OCR system and transcription system are high-quality enough, the errors from candidate generation, such as subtitle annotation error, timestamp inaccuracy, OCR mistake, transcription word error, and text normalization error, are still unavoidable. To further improve the quality of our WenetSpeech corpus, we apply the following validation approach to classify the YouTube OCR and Podcast transcription candidates according to their confidences and filter out the extremely bad candidates.

\vspace{-1em}
\subsubsection{Force Alignment Graph}
\begin{figure}[ht]
  \vspace{-1.5em}
  \centering
  \includegraphics[width=\linewidth]{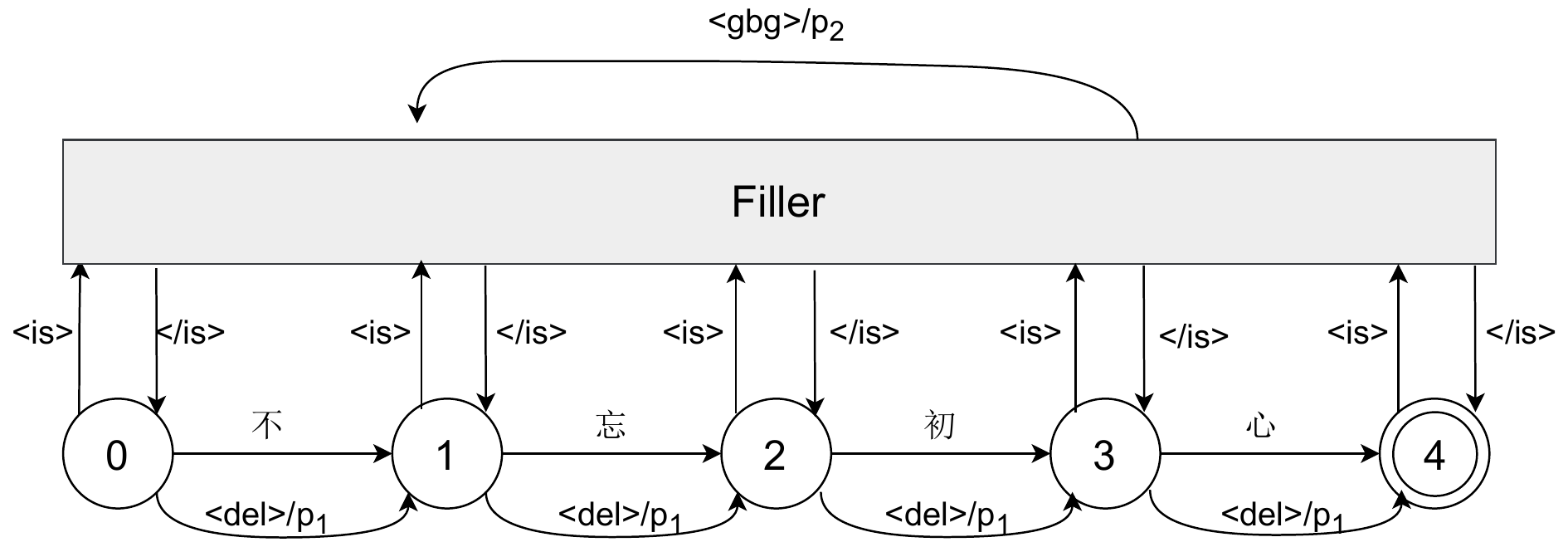}
  \vspace{-2em}
  \caption{An example force alignment graph \textit{L} of "不忘初心"}
  \label{fig:align_with_filler}
  \vspace{-2em}
\end{figure}

\vspace{-0.5em}
Here, we propose an novel CTC-based end-to-end force alignment approach to detect the transcription error.
The transcription is first segmented by the model unit of CTC, and then a \textit{force alignment graph} (\textit{L}) is built for each candidate as shown in Figure \ref{fig:align_with_filler}. The key features of the alignment graph are:

\vspace{-0.5em}
\begin{enumerate}
    \item The oracle transcription alignment path is included.
    \vspace{-0.2em}
    \item Arbitrary deletion operation at any position is allowed through tag \textlangle{}del\textrangle{} with penalty $p_1$.
    \vspace{-0.2em}
    \item Arbitrary insertion or substitution at any position is allowed. From each reasonable position, a start tag \textlangle{}is\textrangle{} and a end tag \textlangle{}/is\textrangle{} are connected to a global \textit{filler} state. On this \textit{filler} state, each CTC modeling unit has a corresponding self-loop arc with penalty $p_2$, which is presented by tag \textlangle{}gbg\textrangle{}.
\end{enumerate}
\vspace{-0.5em}

This makes it possible to capture the error between the audio and the corresponding transcription through decoding technique.

Compared with traditional hybrid validation approach which is used in Librispeech, our proposed approach is a pure end-to-end approach. There is no need for HMM typologies, lexicon, language model components, or careful design of the filler state. So the proposed novel approach simplifies the whole pipeline a lot. The force
alignment graph is implemented by the WeNet toolkit, and it is publicly available \footnote{https:github.com/wenet-e2e/wenet/blob/main/runtime/core/bin/\\label\_checker\_main.cc}.

\begin{CJK*}{UTF8}{gbsn}

\end{CJK*}

\vspace{-0.5em}
\subsubsection{Label Error Detection}
\label{sec:label_error_detection}
\vspace{-0.5em}

After defining the force alignment graph \textit{L}, it is further composed with the CTC topology graph \textit{T}~\cite{miao2015eesen} to build the final \textit{force decoding graph} $\textit{F} = \textit{T} \circ \textit{L}$ for label error detection and validation.

In addition, we assign confidence for each candidate by it's reference (ref) and the force decoding hypothesis (hyp). The confidence $c$ is computed as
\vspace{-1em}
$$
c = 1 - \frac{\text{EditDistance}(ref, hyp)}{\max(\text{len}(ref), \text{len}(hyp))}.
\vspace{-0.5em}
$$

\begin{figure}[ht]
  \vspace{-1em}
  \centering
  \includegraphics[width=\linewidth]{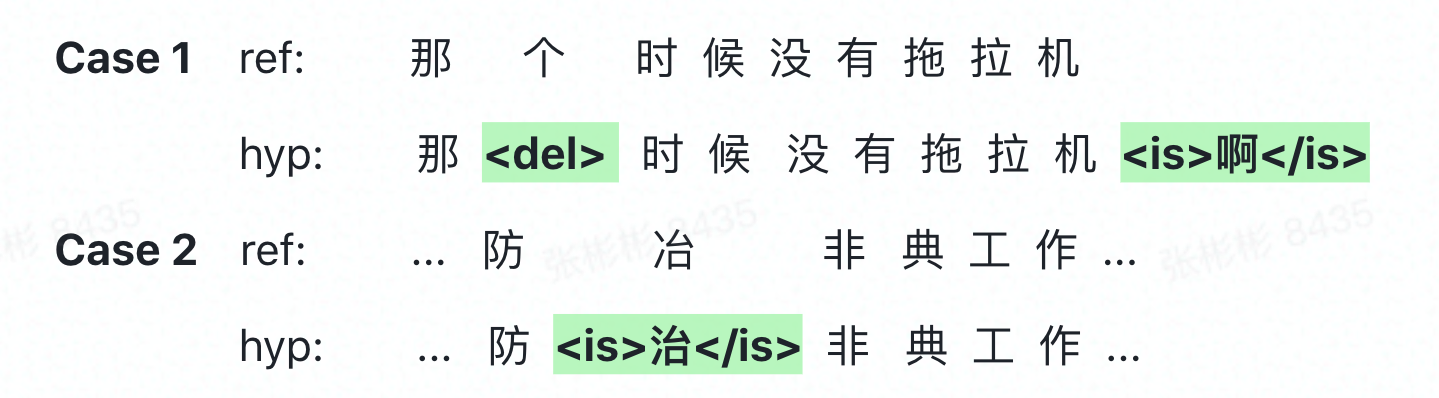}
  \vspace{-2em}
  \caption{Examples of label error detection}
  \label{fig:align_cases}
  \vspace{-1em}
\end{figure}

With the confidence, we classify the audio/text segmentation candidates of our WenetSpeech corpus into \textit{Strong Label} and \textit{Weak Label} sets and even filter out the extremely ones to \textit{Others} set. Figure~\ref{fig:align_cases} shows two real examples that we find by applying label error detection. In \textit{Case 1}, human subtitle error was successfully detected. In \textit{Case 2}, OCR error was successfully detected.

Please note the model used in the transcription system and force alignment system are developed and trained with different model methods and data.
$p_1$ is set to 2.3 and $p_2$ is set to 4.6 in our pipeline.

\vspace{-1.3em}
\section{The WenetSpeech Corpus}\label{sec:corpus}
\vspace{-1em}
In this section, we describe the metadata, audio format, partition by confidence, data diversity, and training and evaluation set design of our WenetSpeech corpus. Instructions and scripts are available at WenetSpeech GitHub repo\footnote{https://github.com/wenet-e2e/WenetSpeech}. 

\vspace{-1.2em}
\subsection{Metadata and Audio Format}
\vspace{-0.5em}

We save all the metadata information to a single JSON file. Local path, original public URL, domain tags, md5, and segments are provided for each audio. And timestamp, label, confidence, subset information are provided for each segment. The design is extensible, and we are planning to add more diverse data in the future.

The original audio files are downloaded and converted to 16k sampling rate, single-channel, and 16-bit signed-integer format. Then Opus compression is applied at an output bit rate of 32 kbps to reduce the size of the WenetSpeech corpus.

\vspace{-1.2em}
\subsection{Size and Confidence}
\vspace{-0.5em}

\begin{table}[t]
\caption{WenetSpeech partition}
\label{tab:confidence}
\centering
\begin{tabular}{@{}lcll@{}}
\toprule[2pt]
Set          & Confidence    & Hours    \\
\midrule[1pt]
Strong Label & [0.95, 1.00]  & 10005    \\
Weak Label   & [0.60, 0.95)  & 2478     \\
Others       & /             & 9952     \\
\midrule[1pt]
Total(hrs)       & /             & 22435 \\ \bottomrule[2pt]
\end{tabular}
\vspace{-1.5em}
\end{table}

We assign confidence for each valid segment which measures the label quality, where the confidence in defined in Section \ref{sec:label_error_detection}.
As shown in Table \ref{tab:confidence}, we select 10005 hours \textit{Strong Label} data, whose confidence is greater than 0.95, as the supervised training data. The 2478 hours \textit{Weak Label} data, whose confidence is between 0.60 and 0.95, is reserved in our metadata for semi-supervised or other usage. At last, \textit{Others} represent all invalid data (i.e. the confidence of data is less than 0.6 or unrecognized) for speech recognition task. In summary, WenetSpeech has 22435 hours of raw audio.

\vspace{-1.2em}
\subsection{Training Data Diversity and Subsets}
\vspace{-0.5em}
We tag all the training data with its source and domain.
All of the training data is from \textit{Youtube} and \textit{Podcast}. As shown in Table \ref{tab:source_domain}, 
we classify the data into 10 groups according to its category.
Please note about 4k hours of data is from drama, which is a special domain with a wide range of themes, topics and scenarios, which may cover any kind of other categories.

\begin{table}[t]
\caption{Training data in different domains with duration (hrs)}
\label{tab:source_domain}
\centering
\begin{tabular}{@{}llll@{}}
\toprule[2pt]
Domain      & Youtube & Podcast & Total   \\ \midrule[1pt]
audiobook   & 0       & 250.9   & 250.9   \\
commentary  & 112.6   & 135.7   & 248.3   \\
documentary & 386.7   & 90.5    & 477.2   \\
drama       & 4338.2  & 0       & 4338.2  \\
interview   & 324.2   & 614     & 938.2   \\
news        & 0       & 868     & 868     \\
reading     & 0       & 1110.2  & 1110.2  \\
talk        & 204     & 90.7    & 294.7   \\
variety     & 603.3   & 224.5   & 827.8   \\
others      & 144     & 507.5   & 651.5   \\ \midrule[1pt]
Total       & 6113  & 3892    & 10005 \\
\bottomrule[2pt]
\end{tabular}
\vspace{-1.5em}
\end{table}

As shown in Table \ref{tab:training_subsets}, we provide 3 training subsets, namely \textit{S, M} and \textit{L} for building ASR systems on different data scales.
Subsets \textit{S} and \textit{M} are sampled from all the training data which have the oracle confidence 1.0. 

\vspace{-0.5em}
\begin{table}[t]
\caption{The training data subsets}
\label{tab:training_subsets}
\centering
\begin{tabular}{@{}llr@{}}
\toprule[2pt]
Training Subsets & Confidence  & Hours \\
\midrule[1pt]
\textit{L}       & [0.95, 1.0] & 10005 \\
\textit{M}       & 1.0         & 1000  \\
\textit{S}       & 1.0         & 100   \\ \bottomrule[2pt]
\end{tabular}
\vspace{-1.5em}
\end{table}

\vspace{-1.1em}
\subsection{Evaluation Sets}
\vspace{-0.5em}
We will release the following evaluation datasets associated with WenetSpeech, and the major information is summarized in Table~\ref{tab:evaluation_sets}.
\begin{enumerate}
    \setlength{\itemsep}{0pt}
    \vspace{-0.5em}
    \item Dev, is specifically designed dataset for some speech tools which require cross-validation in training.
    \vspace{-0.2em}
    \item Test\_Net, is a \textit{matched} test set from the internet. Compared with the training data, it also covers several popular and difficult domains like game commentary, live commerce, etc.
    \vspace{-0.3em}
    \item Test\_Meeting, is a \textit{mismatched} test set since it is a far-field, conversational, spontaneous, and meeting speech dataset. It is sampled from 197 real meetings in a variety of rooms. Its topics cover education, real estate, finance, house and home, technology, interview and so on.
\vspace{-0.5em}
\end{enumerate}
The three evaluation sets are carefully checked by professional annotators to ensure the transcription quality.

\begin{table}[t]
\caption{The WenetSpeech evaluation sets}
\label{tab:evaluation_sets}
\centering
\begin{tabular}{@{}lll@{}}
\toprule[2pt]
Evaluation Sets & Hours & Source       \\
\midrule[1pt]
Dev             & 20    & Internet     \\
Test\_Net       & 23    & Internet     \\
Test\_Meeting   & 15    & Real meeting \\ \bottomrule[2pt]
\end{tabular}
\vspace{-1.5em}
\end{table}

\vspace{-0.8em}
\section{Experiments}\label{sec:exp}
\vspace{-1em}
In this section, we introduce the baseline systems and experimental results on three popular speech recognition toolkits, Kaldi~\cite{povey2011kaldi}, ESPnet~\cite{watanabe2018espnet} and WeNet~\cite{yao2021wenet}.

\vspace{-1em}
\subsection[Kaldi Benchmark]{Kaldi Benchmark\footnote{https://github.com/wenet-e2e/WenetSpeech/tree/main/toolkits/kaldi}}
\vspace{-0.5em}
The Kaldi baseline implements a classical chain model~\cite{povey2016purely} using various amounts of WeNetSpeech data (i.e. S, M, L). We choose the open source vocabulary, BigCiDian\footnote{https://github.com/speechio/BigCiDian}, as our lexicon. And we segment our transcriptions with the open source word segmentation toolkit, jieba~\cite{sun2012jieba}. First, we train a GMM-HMM model to obtain the training alignments. Then, we train a chain model, which stacks 6 convolutional neural network (CNN) layers, 9 factored time-delay neural network (TDNN-F) layers~\cite{povey2018semi}, 1 time-restricted attention layer~\cite{povey2018time} ($H = 30$), 2 projected long short-term memory (LSTMP) with TDNN-F blocks. We feed the 40 dimensional filterbank (FBank) features and 100 dimensional i-vector features as the input. In order to be consistent with the other systems, we only use SpecAugment~\cite{park2019specaugment} technique and abandon the speed/volume perturbation techniques. The chain model is trained by LF-MMI criterion with cross-entropy loss (10 epochs for subset \textit{S}, and 4 epochs for subset \textit{M} and \textit{L}). A 3-gram language model (LM) is used for decoding and generating the lattice. Finally, a recurrent neural network LM (RNNLM) is further adopted to rescore the lattices. The 3-gram and RNNLM are both trained on all the WenetSpeech transcriptions.

\vspace{-1em}
\subsection[ESPnet Benchmark]{ESPnet Benchmark\footnote{https://github.com/wenet-e2e/WenetSpeech/tree/main/toolkits/espnet}}
\vspace{-0.5em}
The ESPnet baseline employs a Conformer model~\cite{gulati2020conformer,guo2021recent} which is designed to capture the global context with the multi-head self-attention module and learn the local correlations synchronously with the convolution module. Our Conformer model consists of a 12-block Conformer encoder ($d^{\text{ff}} = 2048, H = 8, d^{\text{att}} = 512, \text{CNN}_{\text{kernel}} = 15$) and a 6-block Transformer~\cite{vaswani2017attention} decoder ($d^{\text{ff}} = 2048, H = 8$). A set of 5535 Mandarin characters and  26 English letters is used as the modeling units. The objective is a logarithmic linear combination of the CTC ($\lambda = 0.3$) and attention objectives. Label smoothing is applied to the attention objective. During data preparation, we generate 80-dimensional FBank features with a 32ms window and a 8ms frame shift. SpecAugment with 2 frequency masks ($F = 30$) and 2 time masks ($T = 40$), and global CMVN technique are used as data pre-processing. During training, we choose the Adam optimizer with the maximum learning rate of $0.0015$. The Noam learning rate scheduler with 30k warm-up steps is used. The model was trained with dynamic batching skill for 30 epochs. At last, the last 10 best checkpoints were averaged to be the final model. For decoding, the ESPNet system follows the joint CTC/Attention beam search strategy~\cite{hori2017joint}.

\vspace{-1em}
\subsection[WeNet Benchmark]{WeNet Benchmark\footnote{https://github.com/wenet-e2e/WenetSpeech/tree/main/toolkits/wenet}}
\vspace{-0.5em}
The WeNet baseline implements a U2 model~\cite{yao2021wenet}, which unifies streaming and non-streaming end-to-end (E2E) speech recognition in a single model. The basic setup of our WeNet model is same as the ESPnet model except the following minor points: 1) We prepare 80-dimensional FBank features with a 25ms window and a 10ms frame shift. SpecAugment with 2 frequency masks ($F = 30$) and 3 time masks ($T = 50$) and global CMVN technique are applied on the top of our features. 2) The max trainable epoch is 25. Models of the last 10 epochs were averaged to be the final model. The key difference between WeNet and ESPNet is different decoding strategies. Specifically, different from ESPNet's auto-regressive decoding strategy, Wenet generates the N-Best hypotheses by the CTC branch and rescores them by the attention branch.

\vspace{-15pt}
\begin{table}[ht]
\caption{Results (MER\%) on different test sets for baseline systems trained using WenetSpeech training subset L}
\label{tab5:all_baselines}
\centering
\begin{tabular}{@{}lcccc@{}}
\toprule[2pt]
Toolkit & Dev & Test\_Net & Test\_Meeting & AIShell-1 \\
\midrule[1pt]
Kaldi & 9.07 & 12.83 & 24.72 & 5.41 \\
ESPNet & 9.70 & 8.90 & 15.90 & 3.90 \\
WeNet & 8.88 & 9.70 & 15.59 & 4.61 \\
\bottomrule[2pt]
\end{tabular}
\vspace{-10pt}
\end{table}

\vspace{-15pt}
\begin{table}[ht]
\caption{Kaldi baseline results  (MER\%) for different WenetSpeech training subsets}
\label{tab6:kaldi_baselines}
\centering
\begin{tabular}{@{}lcccc@{}}
\toprule[2pt]
SubSet & Dev & Test\_Net & Test\_Meeting & AIShell-1 \\
\midrule[1pt]
L & 9.07 & 12.83 & 24.72 & 5.41 \\
M & 9.81 & 14.19 & 28.22 & 5.93 \\
S & 11.70 & 17.47 & 37.27 & 7.66 \\
\bottomrule[2pt]
\end{tabular}
\vspace{-10pt}
\end{table}

\vspace{-1em}
\subsection {Experimental Results}
\vspace{-0.5em}
We must announce that the results listed here are purely for the purpose of providing a baseline system for each toolkit. They might not reflect the state-of-the-art performance of each toolkit.

In Table~\ref{tab5:all_baselines}, we report the experimental results in Mixture Error Rate (MER)~\cite{shi2020asru}, which considers Mandarin characters and English words as the tokens in the edit distance calculation, on three designed test sets and one well-known, publicly available test set (i.e. AIShell-1~\cite{bu2017aishell} test) with Kaldi, ESPNet and WeNet toolkits respectively. The good performance on AIShell-1 reflects the diversity and reliability of our WenetSpeech corpus. And the results on our designed test sets reflect they are quite challenging. In Table~\ref{tab6:kaldi_baselines}, we provide the Kaldi baseline results for difference scale WenetSpeech subsets. As the growth of the data amount, the performance goes up steadily. 

\vspace{-1em}
\section{Acknowledgements}
\vspace{-1em}
We thank Jiayu Du and Guoguo Chen for their suggestions on this work.
We thank Tencent Ethereal Audio Lab and Xi'an Future AI Innovation Center for providing hosting service for \textit{WenetSpeech}. We also thank MindSpore for the support of this work, which is a new deep learning computing framwork\footnote{https://www.mindspore.cn/}.
Our gratitude goes to Lianhui Zhang and Yu Mao for collecting some of the YouTube data.

\newpage
\scriptsize
\bibliographystyle{IEEEbib}

\newlength{\bibitemsep}\setlength{\bibitemsep}{.15\baselineskip plus .05\baselineskip minus .05\baselineskip}
\newlength{\bibparskip}\setlength{\bibparskip}{0pt}
\let\oldthebibliography\thebibliography
\renewcommand\thebibliography[1]{%
  \oldthebibliography{#1}%
  \setlength{\parskip}{\bibitemsep}%
  \setlength{\itemsep}{\bibparskip}%
}

\bibliography{main}
\end{document}